\documentclass[submission,copyright,creativecommons]{eptcs}
\providecommand{\event}{WORDS 2011} 
\usepackage{breakurl}             

\usepackage{array}
\usepackage{amsmath}
\usepackage{amsfonts}
\usepackage{amssymb}
\usepackage{amsthm}
\usepackage{enumerate}
\usepackage{algorithm}
\usepackage{floatflt}
\usepackage[footnotesize]{caption}
\usepackage[latin1]{inputenc}
\usepackage{tikz}
\usetikzlibrary{arrows}
\usepackage{calc}

\renewcommand{\hat}{\widehat}

\newcommand{\Freeman}{\mathcal{F}}

\newcommand{\N}{\mathbb{N}}
\newcommand{\Z}{\mathbb{Z}}

\newcommand{\Reals}{{\mathbb R}}

\newcommand{\ConvHull}{\ensuremath{\mathrm{Conv}}}
\newcommand{\UpConvHull}{\ensuremath{\mathrm{Conv^+}}}
\newcommand{\LoConvHull}{\ensuremath{\mathrm{Conv^-}}}
\newcommand{\Pm}[2]{\Pmm{#1}(#2)}
\newcommand{\Pmm}[1]{\ensuremath{\phi_{#1}}}

\newcommand{\gras}[1]{{\bf #1}}
\newcommand{\0}{\gras{0}}
\newcommand{\1}{\gras{1}}
\newcommand{\2}{\gras{2}}
\newcommand{\3}{\gras{3}}

\newcommand{\Alphabet}{\mathcal{A}}

\newcommand{\Supp}{\mathrm{Alph}}
\newcommand{\G}{{\mathcal G}}
\newcommand{\Fact}{\mathrm{Fact}}
\newcommand{\Pref}{\mathrm{Pref}}

\newcommand{\Pal}{\mathrm{Pal}}

\newcommand{\LPS}{\hbox{\rm LPS}}

\newcommand{\bse}{\ensuremath{{\boldsymbol{\epsilon}}}}

\newcommand{\til}{\widetilde}
\newcommand\txtc[1]{\text{\textcircled{{$#1$}}}}
\newcommand\boldit[1]{\textbf{\emph{#1}}}
\newcommand{\Base}{\ensuremath{\mathbb{B}}}

\newcommand{\bx}{\ensuremath{{\boldit{x}}}}
\newcommand{\by}{\ensuremath{{\boldit{y}}}}

\newcommand{\Cancel}{\mathcal{C}}
\newcommand{\Forbidden}{\ensuremath{\{\0\2,\2\0,\1\3,\3\1\}}}

\newcommand{\ligne}[1]{\makebox[15pt]{\hfill$#1:$}}

\newcommand{\ind}[3]{\\ \vspace{-1.3mm} \ligne{#1}\hspace{2mm}\hspace{#2mm}\hspace{#2mm}\hspace{#2mm} \textrm{\emph{#3}} \smallskip }

\newsavebox{\gauche}
\sbox{\gauche}{\setlength{\unitlength}{1mm}
                  \begin{picture}(1.5,3)(0,0)
                  \thicklines
                  {\put(0,-.5){\line(1,0){1}}}%
                  {\put(0.15,-.64){\line(0,3){3.48}}}%
                  {\put(0,2.7){\line(1,0){1}}}%
                  \end{picture}}
\newcommand{\lbr}{\usebox{\gauche}}
\newsavebox{\droite}
\sbox{\droite}{\setlength{\unitlength}{1mm}
                  \begin{picture}(1,3)(0,0)
                  \thicklines
                  {\put(0,-.5){\line(-1,0){1}}}%
                  {\put(-0.16,-.64){\line(0,3){3.48}}}%
                  {\put(0,2.7){\line(-1,0){1}}}%
                  \end{picture}}
\newcommand{\rbr}{\usebox{\droite}}
\newcommand\Class[1]{\hspace{-1pt}\lbr#1\rbr\hspace{-1pt}}

\newsavebox{\TN}
\sbox{\TN}{\setlength{\unitlength}{1mm}
                  \begin{picture}(2,2)(0,0)
                  \thinlines
                  {\put(0,-0.3){\large$\mathcal T$}}%
                  {\put(1.7,1){\circle{1.6}}}%
                  \end{picture}}
\newcommand{\TNr}{\usebox{\TN\;}}

\newsavebox{\ODLT}
\sbox{\ODLT}{\setlength{\unitlength}{1mm}
                  \begin{picture}(2,2)(0,0)
                  \thinlines
                  {\put(-1,-0.2){\large$\mathcal D$}}%
                  {\put(0.1,1.16){\circle{1.6}}}%
                  \end{picture}}

\newcommand{\Turns}{\ensuremath{\mathcal{T}}}
\newcommand{\oTurns}{\TNr}

  {\end{enumerate}}

\newtheorem{theorem}{Theorem}
\newtheorem{definition}{Definition}
\newtheorem{proposition}[theorem]{Proposition}

\newtheorem{lemma}[theorem]{Lemma}

\begin{document}

\title{Interactions between Digital Geometry and\\  Combinatorics on Words}

\def\titlerunning{Interactions between Digital Geometry and Combinatorics on Words}
\def\authorrunning{S. Brlek}

\author{Sre\v{c}ko Brlek
\institute{LaCIM, Universit{\'e} du Qu{\'e}bec {\`a} Montr{\'e}al,\\
C. P. 8888 Succursale ``Centre-Ville'', Montr{\'e}al (QC), CANADA H3C 3P8}
\email{brlek.srecko@uqam.ca}}

\maketitle

\begin{abstract}
We review some recent results  in digital geometry obtained by using a combinatorics on words approach to discrete geometry. Motivated on the one hand by the well-known theory of Sturmian words which model conveniently discrete lines in the plane, and on the other hand by the development of digital geometry, this study reveals strong links between the two fields. Discrete figures are identified with polyominoes encoded by words. The combinatorial tools lead to elegant descriptions of geometrical features and efficient algorithms. Among these, radix-trees are useful for  efficiently detecting path intersection, Lyndon  and Christoffel words appear as the main tools for describing digital convexity; equations on words allow to better understand tilings by translations. 
\end{abstract}

\section{Introduction}

The expansion of computers has led in the last few decades to several breakthrough in technological achievements. Among these, digital imaging is increasing its spread  and is extensively used in a wide range of applications such as image synthesis,  remote sensing, medical image processing to cite a few. Developed mostly by the engineering world, digital geometry has led to the discovery (sometimes a rediscovery) of new results about discrete sets, concurrently to the design of new algorithms tools,  and enriched the  broader field of discrete geometry.\smallskip

Combinatorics on words has imposed itself as a powerful tool for the study  of  discrete, linear, and non-commutative objects that appear in almost any branches of mathematics, and discrete geometry is not an exception.  Traditionally, digital geometry works on characterization and recognition of discrete objects using an arithmetic approach or computational geometry. However combinatorics on words provide some  useful tools and efficient algorithms for handling discrete objects. 
Lothaire's books \cite{lothaire1,lothaire2,lothaire3} constitute the  reference for  presenting a unified view on combinatorics on words and many of its applications.\\

 As mentioned by Klette and Rozenfeld  in their survey on digital straightness \cite{KleRo}
\begin{quote}
``{\it  Related work even earlier on the theory of words, specifically, on mechanical or Sturmian words, remained unnoticed in the pattern recognition community}"
\end{quote}
there was a need for new investigations where combinatorics on words 
would enrich the classical Euclidean approach of digital geometry. \smallskip

We revisit some classical problems in discrete geometry from this new point of view. For our purpose the discrete plane is identified with the square grid $\Z\times \Z$.

\section{Preliminaries}
We refer to Lothaire \cite{lothaire1} for the basic terminology and notation about words on a finite alphabet $\Alphabet$. It includes  the {\it empty} word $\varepsilon$, {\it length, , conjugate, factor, prefix, suffix, proper factor,  free monoid $\Alphabet^*$,  morphism, antimorphism, occurrences, palindrome, period, power, primitive, reversal}.
The set of all factors of $w$ is denoted by  $\Fact(w),$  those of
length  $n$ is $\Fact_n(w) = \Fact(w) \cap \Alphabet^n,$ $\Pref(w)$ is
the set of all prefixes of $w$, and  the set of its palindromic factors is $\Pal(L)$.  If $w=pu$, with $|w|=n$ and $|p|=k$, then 
$p^{-1}w = w[k..n-1] = u$ is the word obtained by erasing from $w$ its prefix $p$.
The  class of a word $w$ is denoted $\Class{w}$. Every  word contains palindromes, the letters and $\varepsilon$ being necessarily part of them. This justifies the introduction of the function $\LPS : \Alphabet^* \to \Alphabet^*$ which associates  to any  word $w$  its longest palindromic suffix $\LPS(w)$.
Given a  total order $<$ on $\Alphabet$, the \emph{lexicographic ordering} is defined as usual. 
\paragraph{Lyndon words}
Introduced as \emph{standard lexicographic sequences} by Lyndon in 1954, Lyndon words have several characterizations (see \cite{lothaire1}). We shall define them as words being strictly smaller than any of their circular permutations. 
\begin{definition}
A Lyndon word $l \in \Alphabet^+$ is a word such that $l = uv$ with $u,v \in \Alphabet^+$ implies that $l<vu$. 
\end{definition}

Note that Lyndon words are always primitive. The most important result about Lyndon words is the following unique factorization theorem (see Lothaire \cite{lothaire1} Theorem 5.1.1).
\begin{theorem}  Any word $w\in \Alphabet^+$  admits a unique factorization as a sequence of decreasing Lyndon words:
\begin{equation}\label{factoLyndon}
w = l_1^{n_1} l_2^{n_2} \cdots l_k^{n_k}, \quad l_1 > l_2 > \dots > l_k
\end{equation}
where  $n_i\geq 1$ and $l_i$ is a Lyndon word, for all $i$ such that $1\leq i\leq k$. 
\end{theorem}
There exist several algorithms for factorizing a word $w=w_1w_2\cdots w_n$ into Lyndon words and the more efficient are linear . 
An  elegant one was invented by Duval \cite{Duv}. It works by reading from left to right, with at most $2n$ comparisons of letters (see also \cite{Reu}, Section 7.4). 
Another one, uses the concept of \emph{suffix standardization} of the word $w$, 
and builds a \emph{suffix array} of $w$, which may be computed in linear time \cite{CroHanLec}. Then the Lyndon factorization of $w$ is obtained by cutting $w$ just before each left-to-right minimum of its suffix-array.

\paragraph*{A quadtree with a radix tree structure for points in the integer plane {\rm(\cite{BKPproc,BKP})}}
Let $\Base=\{0,1\}$ be the base  for writing  integers. Words in $\Base^*$ are conveniently represented in the {\em radix order} by a complete binary tree (see for instance \cite{knuth3,lothaire3}), where the level $k$ contains all the binary words of length $k$, and the order is given by the breadth-first traversal of  the tree. To distinguish a natural number $x\in \N$ from its representation we write $\bx\in \Base^*$.  The edges are defined inductively by the rewriting rule $\bx \to \bx\cdot 0 + \bx\cdot 1,$ with the convention that  $0$ and $1$ are the labels of, respectively,  the left and right edges of the node having value $\bx$. This representation  is extended to $\Base^*\times \Base^*$ as follows. As usual, the concatenation is extended to the cartesian product of words by setting for $(\bx,\by) \in \Base^*\times \Base^*$, and $(\alpha,\beta) \in \Base\times \Base$
$$(\bx,\by)\cdot(\alpha,\beta)= (\bx\cdot\alpha,\by\cdot\beta).$$
Let $\bx$ and $\by$ be two binary words having same length.  Then the rule 
\begin{equation}(\bx,\by) \to (\bx\cdot 0, \by\cdot 0) + (\bx\cdot 0, \by\cdot 1)  + (\bx\cdot 1, \by\cdot 0)  +(\bx\cdot 1, \by\cdot 1) \label{TreeRule}
\end{equation}
defines a   $\G' =(N,R)$, sub-graph  of  $\G = (N,R,T)$,   such that :
\begin{enumerate}[\rm (i)]
 \item the root is labeled $(0, 0)$;
 \item each  node (except the root) has four sons;
 \item if a node is labeled $(\bx,\by)$ then $|\bx|=|\by|;$ 
 \item edges are undirected, e.g. may be followed in both directions.
 \end{enumerate}
 By convention, edges leading to the sons have labels from the ordered set $\{(0,0), (0,1), (1,0), (1,1)\}$. These labels equip  the quadtree  with a \emph{radix tree} structure for Equation \eqref{TreeRule} implies that $(x', y')$ is a son of $(x,y)$, if and only if
\[(x', y') = (2x+ \alpha, 2y+\beta),\] 
for some   $(\alpha, \beta) \in  \Base\times \Base$. Observe that any pair $(x, y)$ of nonnegative integers is represented exactly once in this tree. Indeed, if  $|\bx| =|\by|$  (by filling with zeros at the left of the shortest one), the sequence of pairs of digits (the two digits in first place, the two digits in second place, and so on) gives the unique path in the tree leading to this pair. Of course the root may have up to three sons since no edge labeled $(0,0)$  starts from the root.\\ 
\begin{figure}[h!]
\centering
\includegraphics[width=2.75in]{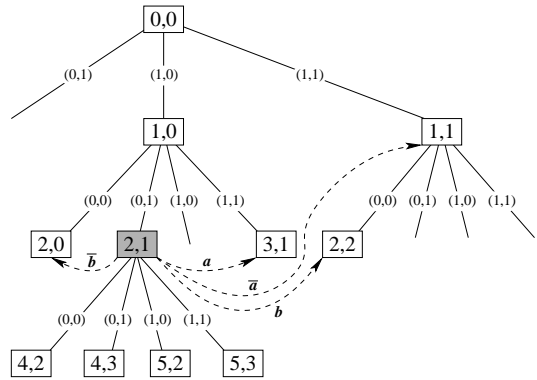}
\caption{The  point $(2,1)$ with its neighbors.}\label{RadixTree}
\end{figure}
\paragraph{Neighboring links {\rm\cite{BKPproc,BKP}}}
Given  $(x,y) \in \Z^2$, a point $(x',y')$ is an {\em \bse-neighbor} of  $(x,y)$ if there exists $\epsilon \in \Freeman$ such that  
$(x',y') = (x,y)+\emph{\bse}=(x+\epsilon_1,y+\epsilon_2).$\smallskip

We superpose  on $G'$  the neighboring relation given by the edges of $T$ (dashed lines). More precisely, for each elementary translation $\epsilon\in\Freeman$, each  node $\txtc{z}=(x,y)$ is linked  to its  $\epsilon$-neighbor  $\txtc{z} +\bse$, when it exists. 
If a level $k$ is fixed, it is easy to construct the graph 
\[\G^{(k)} = (N^{(k)} ,R^{(k)} ,T^{(k)} )\]
 such that 
\begin{enumerate}[\rm (i)]
\item if $(\bx,\by)\in N^{(k)}$, then  $|\bx| =|\by|=k$; 
\item the functions  $N^{(k)} \hookrightarrow  \N\times\N  \hookrightarrow  \Base^*\times\Base^* $ are injective;
\item $R^{(k)}$ is the radix-tree representation :  $(\Base^{<k}\times\Base^{<k})\times  (\Base\times\Base)\stackrel{\bullet}{\to} \Base^{\leq k}\times\Base^{\leq k}$;\smallskip
\item the neighboring relation is $T^{(k)} \subseteq N\times(\Base\times\Base)\times N$.
\end{enumerate}

Note that the labeling in Fig. \ref{RadixTree} is superfluous:  each node represents indeed an integer unambiguously determined by the path from the root using edges in $R$; similarly for the ordered edges.
Moreover, if a given subset $M\subset \N\times\N$ has to be represented, then one may trim the unnecessary nodes so that  the corresponding graph  $\G_M$ is not necessarily complete. \smallskip
 
Recall that adding $1$ to an integer $\bx \in \Base^k$  is easily performed by a sequential function. 
Indeed, every positive integer can be written $\bx = u 1^i 0^j$, where $i\geq 1$,  $j\geq 0$, with 
$u\in \{\varepsilon\} \cup \left\{\Base^{k-i-j-1}\cdot 0\right\}.$ In other words, $1^j$ is the last run of $1$'s.
The piece of code for adding 1 to an integer written in base 2 is\medskip
$
\ind{1}{0}{{\bf If} $j\not = 0 $ {\bf then} Return $u1^i0^{j-1}1;$}
\ind{2}{5}{{\bf else If} $u= \varepsilon$ {\bf then} Return $1\cdot 0^i;$}
\ind{3}{13}{{\bf else}  Return $u\cdot 0^{-1}\cdot 1\cdot 0^i;$}
\ind{4}{8}{{\bf end if}}
\ind{5}{0}{{\bf end if}}
$\medskip

\noindent where $0^{-1}$ means to erase a $0$. Clearly, the computation time of this algorithm  is proportional to the length of the last run of $1$'s.  Much better is achieved with  the radix tree structure, where, given a node $\txtc{z}$, its {\em father} is denoted  $f(\txtc{z})$, and we write $f(x,y)$ or $f(\bx,\by)$ if its label is $(x,y)$.
The following technical lemma is a direct  adaptation to $\Base^*\times\Base^*$  of the addition above.
\begin{lemma}\label{condition} Let $G^{(k)}$ be the complete graph representing $\Base^{\leq k}\times\Base^{\leq k}$ for some $k\geq 1$,  $\epsilon \in \Freeman$, and $\txtc{z} = (\emph{\bx, \by})$ be a node of $N^{k}$. If one of the four conditions holds:
\[\begin{array}{rlcccrlcc}
 \text{\rm(i) } &\epsilon = \0 & \text{\rm and} &  \emph{\bx[k]} = 0, & &
\text{\rm(ii) }  &\epsilon = \2 & \text{\rm and} & \emph{\bx[k]} = 1, \\
\text{\rm(iii) }  &\epsilon = \1 & \text{\rm and} & \emph{\by[k]} = 0, & &
\text{\rm(iv) }  &\epsilon = \3 & \text{\rm and} & \emph{\by[k]} = 1,
\end{array} \]
then $f(\txtc{z}) = f\left(\txtc{z} +\bse\right)$. Otherwise, $f(\txtc{z}) + \bse = f(\txtc{z} + \bse)$.
\end{lemma}
\medskip
\begin{floatingfigure}[h!r]{3in}
\centering
\includegraphics[height=1in]{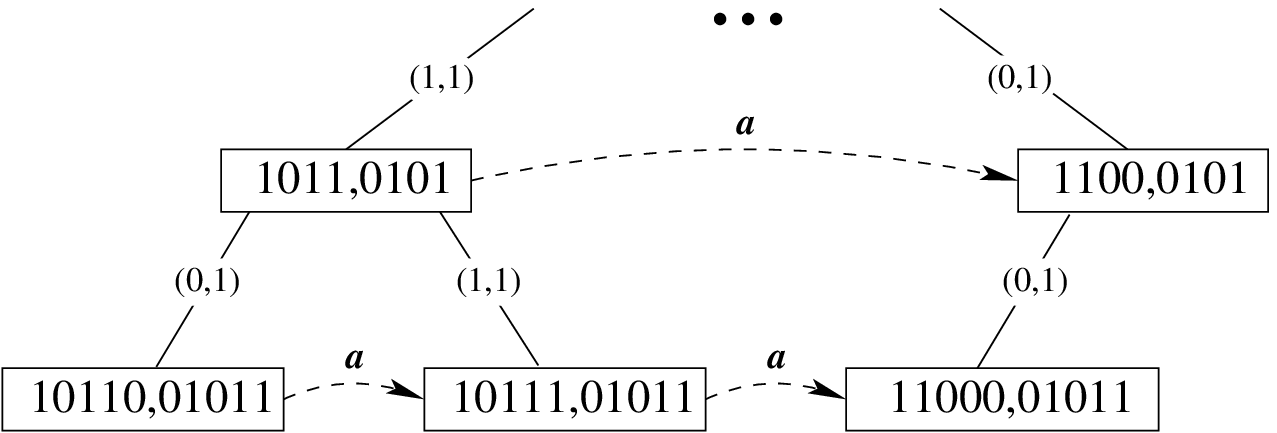}
\end{floatingfigure}
\noindent  The process is illustrated for case (i) in the diagram on the right
where the nodes\vspace{4pt} 
\hbox{$\,(10110, \bullet)~$ and  $\,(10111, \bullet)\,$}\medskip
\\
share the same father while fathers of neighboring nodes\medskip\\
\smallskip
 $(\bullet, 01011)~$ and $~(\bullet, 01011)$\\
\smallskip \hbox{are distinct but share the same neighboring relation.}

\paragraph{A representation for paths in the square grid}

Here, we encode paths with the so-called \emph{Freeman chain code}\cite{freeman1} based on the  alphabet $\Freeman=\{\0,\1,\2,\3\}$, considered as the additive group of integers $\bmod~4$. 
Basic transformations on $\Freeman$ are rotations $\rho^i:x\mapsto x+i$ and reflections $\sigma_i:x\mapsto i - x$,  which extend uniquely to morphisms (w.r.t  concatenation) on $\Freeman^*$. 
Given a nonempty word $w\in\Freeman^*$, the \emph{first differences word} $\Delta(w)\in\Freeman^*$ of $w$ is
\begin{equation}\label{Delta}
\Delta(w) = (w_2-w_1)\cdot (w_3-w_2) \cdots (w_n - w_{n-1}).
\end{equation}
One may verify that if $z \in\Freeman^*$, then $\Delta(wz)=\Delta(w)\Delta (w_nz_1)\Delta(z)$.
Words in $\Freeman^*$ are interpreted as paths in the square grid, so that we indistinctly talk of any word $w\in\Freeman^*$ as the \emph{path} $w$. 
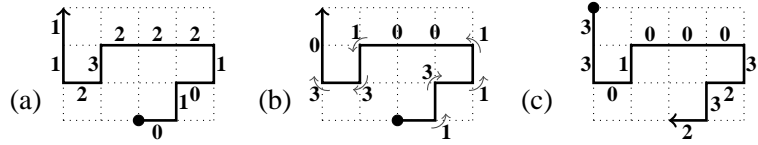
\begin{figure}[ht]
\centering
\begin{tabular}{ccc}


\begin{tikzpicture}[xscale=.5pt, yscale=.5pt,inner sep=0mm,point/.style={circle,draw=black,fill=black, minimum size=4pt}]

\def\ub{ -- ++(0,1)}
\def\ua{ -- ++(1,0)}
\def\uA{ -- ++(-1,0)}
\def\uB{ -- ++(0,-1)}
\def\chemin{\ua\ub\ua\ub\uA\uA\uA\uB\uA\ub\ub}
\draw[step=1cm,black,thin,dotted] (-2,0) grid (2,3);

\draw[black,line width=1,->](0,0)\chemin;
\node at (0,0)[point]{};

\foreach \i in {(0.5,-0.3), (1.5,0.7)}
    \node at \i{\scriptsize{\0}};
\foreach \i in {(1.2,0.5), (2.2,1.5),(-2.2,1.5),(-2.2,2.5)}
    \node at \i{\scriptsize{\1}};
\foreach \i in {(-0.5,2.3),(0.5,2.3),(1.5,2.3),(-1.5,0.7)}
    \node at \i{\scriptsize{\2}};
\foreach \i in {(-1.2,1.5)}
    \node at \i{\scriptsize{\3}};
\node at (-3,0.5){(a)};

\end{tikzpicture}

&

\begin{tikzpicture}[xscale=.5pt, yscale=.5pt,inner sep=0mm,point/.style={circle,draw=black,fill=black, minimum size=4pt}]

\def\ub{ -- ++(0,1)}
\def\ua{ -- ++(1,0)}
\def\uA{ -- ++(-1,0)}
\def\uB{ -- ++(0,-1)}
\def\chemin{\ua\ub\ua\ub\uA\uA\uA\uB\uA\ub\ub}
\draw[step=1cm,black,thin,dotted] (-2,0) grid (2,3);

\draw[black,line width=1,->](0,0)\chemin;
\node at (0,0)[point]{};

\draw[->,color=black!70] (0.9,-0.2) arc (-95:5:10pt);
\draw[->,color=black!70] (0.8,0.8) arc (185:85:10pt);
\draw[->,color=black!70] (1.9,0.8) arc (-95:5:10pt);
\draw[->,color=black!70] (2.2,1.8) arc (-5:95:10pt);
\draw[->,color=black!70] (-.8,2.2) arc (85:185:10pt);
\draw[->,color=black!70] (-0.8,1.2) arc (5:-95:10pt);
\draw[->,color=black!70] (-1.8,.8) arc (280:175:10pt);

\foreach \i in {(1,2.35),(0,2.35),(-2.2,2)}
    \node at \i{\scriptsize{\0}};
    
\foreach \i in {(1.3,-.3),(2.3,0.7),(2.3,2.3),(-1.1,2.35)}
    \node at \i{\scriptsize{\1}};
    
\foreach \i in {(0.8,1.3),(-.8,.7),(-2.2,0.7)}
    \node at \i{\scriptsize{\3}};
\node at (-3.25,0.5){(b)};

\end{tikzpicture}

&

\begin{tikzpicture}[xscale=.5pt, yscale=.5pt,inner sep=0mm,point/.style={circle,draw=black,fill=black, minimum size=4pt}]

\def\ub{ -- ++(0,1)}
\def\ua{ -- ++(1,0)}
\def\uA{ -- ++(-1,0)}
\def\uB{ -- ++(0,-1)}
\def\chemin{\ua\ub\ua\ub\uA\uA\uA\uB\uA\ub\ub}
\draw[step=1cm,black,thin,dotted] (-2,0) grid (2,3);

\draw[black,line width=1,<-](0,0)\chemin;
\node at (-2,3)[point]{};

\foreach \i in {(0.5,-0.3), (1.6,0.7)}
    \node at \i{\scriptsize{\2}};
\foreach \i in {(1.25,0.45), (2.2,1.5),(-2.2,1.5),(-2.2,2.5)}
    \node at \i{\scriptsize{\3}};
\foreach \i in {(-0.5,2.3),(0.5,2.3),(1.5,2.3),(-1.5,0.7)}
    \node at \i{\scriptsize{\0}};
\foreach \i in {(-1.2,1.5)}
    \node at \i{\scriptsize{\1}};

\node at (-3.5,0.5){(c)};
\end{tikzpicture}

\end{tabular}
\caption{(a)  $w=\gras{01012223211}$. (b)  $\Delta(w) = \gras{1311001330}$. (c)  $\hat{w}=\gras{33010003232}$.}
\label{figdelta}
\end{figure}
Moreover, the word $\hat{w}:=\rho^2(\widetilde{w})$ is \emph{homologous} to $w$, i.e.,   in direction opposite to that of $w$ (Figure~\ref{figdelta}). 
%
A word $u\in\Freeman^*$ may contain factors in $\Cancel=\Forbidden$, corresponding to cancelling steps on a path.  Nevertheless, each word $w$ can be reduced in a unique way to a word $w'$,  by sequentially applying the rewriting rules in $\{u\mapsto \varepsilon \mid u \in \Cancel\}$. The \emph{reduced word}  $w'$ of $w$ is nothing but a word in $\mathcal P=\Freeman^*\setminus\Freeman^*\Cancel\Freeman^*$.
The \emph{turning number}\footnotemark[1] of $w$ is defined by $\Turns(w)=\left(|\Delta(w')|_\1 -|\Delta(w')|_\3 \right)/4$.
\footnotetext[1]{In \cite{bll2,bll3}, the authors introduced the notion of  \emph{winding number}  of $w$ which is $4\Turns(w)$.}\medskip

A path $w$ is \emph{closed}  if it satisfies $|w|_\0 = |w|_{\2}$ and $|w|_\1 = |w|_{\3}$, and it  is \emph{simple} if no proper factor of $w$ is  closed. 
A \emph{boundary word} is a simple and closed path, and a \emph{polyomino} is a subset of $\Z^2$ contained in some boundary word.  It is convenient to represent each closed path $w$  by its conjugacy class $\Class{w}$, also called \emph{circular word}. An adjustment is necessary to the function $\Turns$, for we take into account the closing turn. The first differences also noted $\Delta$ is defined on any closed path $w$ by setting
\[\Delta(\Class{w})\equiv \Delta(w) \cdot (w_1 - w_n), \]
which is also a closed word.
By applying the same rewriting rules, a circular word $\Class{w}$ is \emph{circularly-reduced} to a unique  word $\Class{w'}$. If $w$ is a closed path, then the \emph{turning number}\footnotemark[1] of $w$ is 
\[\oTurns(w)=\Turns(\Class{w})=\left(|\Delta(\Class{w'})|_\1 -|\Delta(\Class{w'})|_\3 \right)/4.\]
 It corresponds to its total curvature divided by $2\pi$. Clearly, the turning number $\Turns(\Class{w})$ of a closed path $w$ belongs to $\Z$  (see \cite{bll2,bll3}).

\paragraph{The convex hull of a finite set of points}The lexicographic order $<$ on points of $\Reals^2$ or $\Z^2$ is such
 that $(x,y) < (x',y')$ when either $x < x'$ or $x=x'$ and $y<
 y'$. The {\em convex hull} of a finite set $S$ of points in $\Reals^2$
 is the intersection of all
 convex sets containing these points and is denoted by
 $\ConvHull(S)$. $S$ being finite, it is clearly a polygon in the
 plane whose vertices are elements of $S$. The {\em upper convex
 hull} of $S$, denoted by $\UpConvHull(S)$, is the clockwise oriented
 sequence of consecutive edges of $\ConvHull(S)$ starting from the
 lowest vertex and ending on the highest vertex. The {\em lower
 convex hull} of $S$, denoted by $\LoConvHull(S)$, is the clockwise
 oriented sequence of consecutive edges of $\ConvHull(S)$ starting
 from the highest vertex and ending on the lowest vertex.

\section{The Daurat-Nivat relation {\rm \cite{bll5,bll6}}}

 We recall from  Daurat and Nivat \cite{dauniv} that a {\em discrete} set $E$ is a subset of $Z^{2}$ and an element $(i,j) \in E$ corresponds to a unit square with vertices $(i \pm \frac{1}{2}, j \pm \frac{1}{2})$. One sets $P_{1/2}=(\frac{1}{2},\frac{1}{2})+ Z^{2}$ and the salient and reentrant points are defined as follows:\medskip
\begin{definition} {\rm (Daurat and Nivat\cite{dauniv})} Let $E$ be a discrete set. Then 
\begin{itemize}
\item[\rm(i)]A {\em corner} is a couple $({\bf M},{\bf N})$ when ${\bf M} \in P_{1/2}$ and ${\bf N} \in Z^{2}$ and ${\bf M}-{\bf N}$ is in $(\{ \pm \frac{1}{2}, \pm \frac{1}{2} \})$.
\item[\rm(ii)] A corner $({\bf M},{\bf N})$ is  {\em salient } if ${\bf N} \in E$ and ${\bf M}$ is the extremity of two edges of the border of $E$ which are also sides of the square ${\bf N}+ [ -\frac{1}{2},\frac{1}{2} ]^{2}$.
\item[\rm(iii)] A corner $({\bf M},{\bf N})$ is  {\em reentrant } if ${\bf N} \in E$ and ${\bf M}$ is the extremities of two consecutive edges of the border $E$ which are not sides of the square ${\bf N}+ [- \frac{1}{2},\frac{1}{2} ]^{2}$.
\end{itemize}
\end{definition} 
\begin{definition}{\rm (Daurat and Nivat\cite{dauniv})}
The multiset of border-salient (resp. border-reentrant) points of a discrete set $E$, denoted $S_{B}(E)$ (resp. $R_{B}(E)$) is the multiset whose support is included in $P_{1/2}$ and such that for any ${\bf M} \in P_{1/2}$, the number ${\rm mult}_{S_{B}(E)}({\bf M})$ (resp.  ${\rm mult}_{R_{B}(E)}({\bf M})$) is the number of ${\bf N}$ such that $({\bf M},{\bf N})$ is a salient corner (resp. reentrant corner). 
\end{definition}

\begin{figure}[h]
\begin{center}
\includegraphics[width=0.25\textwidth]{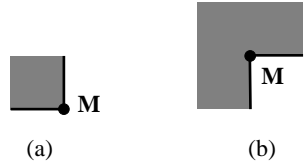}
\caption{Salient and reentrant points in the terminology of Daurat and Nivat}\label{SRpoints}
\end{center}
\end{figure}

In other words, in their terminology modulo a translation by $(\frac{1}{2}, \frac{1}{2})$, a point ${\bf M}$ on the boundary of a polyomino $P$ is  salient (see Figure \ref{SRpoints} (a)) if it belongs to the intersection of two consecutive sides of a square belonging to $P$. The point ${\bf M}$ is  reentrant  (see Figure \ref{SRpoints} (b)) if it is the intersection of two consecutive edges of the contour of $P$ which are sides of a square not belonging to $P$.

In our framework, a salient point of the boundary $w$ of a polyomino corresponds to a left turn (a $\1$ in $\Delta(w)$) and a reentrant one to a right one (a $\3$ in $\Delta(w)$), provided the traversal is done in a counterclockwise manner. The Daurat-Nivat \cite{dauniv} states that the $S$ salient and $R$ reentrant points in every polyomino are related by the formula
\begin{equation}\label{SR} S - R = 4. 
\end{equation}

\begin{figure}[h!]
\begin{center}
\includegraphics[width=0.49\textwidth]{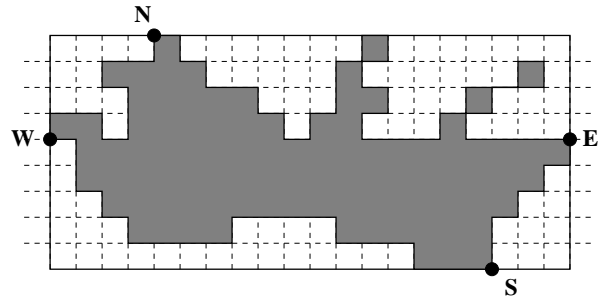}
\caption{A  region and its four extremal points.}\label{region}
\end{center}
\end{figure}
The four {\it extremal} points are defined by the coordinates: $\bf W$  is the lowest intersection with the left side of the bounding rectangle $Q$, $\bf N$ the leftmost intersection with the top side, $\bf E$ the highest intersection with the right side,   and $\bf S$ the rightmost intersection with the bottom side. 
Note that the four extra salient points can be canonically identified as 
$\bf W$, $\bf S$, $\bf E$ and ${\bf N}$.
Then, the Daurat-Nivat relation \cite{dauniv} may be restated as 
\begin{proposition}\label{turning_bw}
The turning number of a boundary word $w$ is $\oTurns(w)=\pm 1$.
\end{proposition}

Note that  a boundary word $w$ is \emph{positively oriented} (counterclockwise) iff its turning number is $\oTurns(w)=1$. 

\paragraph{Remarks} 1. This rather elementary geometrical property is fundamental in proving a result about deciding whether a polyomino tiles the plane by translation or not. 

2. This result can be extended easily to take into account hexagonal grids in which case the alphabet $\Freeman$ must be extended to 6 letters, in which case the Daurat-Nivat relation becomes (see \cite{bll5,bll6} for more details) 
\begin{equation}\label{SR6} S - R = 6. 
\end{equation}

3. The statement above includes closed paths not necessarily simple. It cannot therefore be used to determine whether a closed path is simple or not.

\section{Path Intersection {\rm \cite{BKPproc,BKP}}}

Many problems in discrete geometry involve the analysis of the contour of discrete sets  and many  problems are solved by using linear algorithms in the length of the contour word. However, most of the time it is assumed that the path encoded by this word does not intersect itself. Checking non intersection  amounts to check if a grid point is visited twice. Of course, one might easily provide an ${\cal O}(n \log n)$ algorithm where sorting is involved, or use hash tables providing a linear time algorithm on average but not in worst case.  
%
The underlying principle of the algorithm is to build a graph  $\G = (N,R,T)$ where $N$ is a set of nodes associated to points of the plane, $R$ and $T$ are two distincts sets of oriented edges. The edges in $R$ give a quadtree structure on the nodes while the edges in $T$ are links from each node to its neighbors.

\paragraph{The Algorithm}\label{secAlgo}

First, we assume that  the path is coded by a word $w$ starting at the origin $(0,0)$, and stays in the first quadrant $\N\times \N$. This means that the coordinates of all points are nonnegative. Note that in $\N\times \N$, each point has exactly four neighbors with the exception of the origin $(0,0)$ which admits only two neighbors, namely $(0,1)$ and $(1,0)$,  and the points on the half lines $(x,0)$ and $(0,y)$ with $x,y\geq 1$ which admit only three neighbors.

Now, assume that  the node  $(\bx,\by)$ exists and that  its neighbor $(x+1,y+0)$ does not. If $|\bx|=|\by|=k$, then the translation $(x,y)+(1,0)$ is obtained in three steps by the following rules: 
\begin{enumerate}
\item take the edge in $R$ to $f(\bx,\by)=(\bx[1..k-1], \by[1..k-1])$;
\item take (or create) the edge in $T$ from  $f(\bx,\by)$ to $\txtc{z}=f(\bx,\by)+(1, 0)$; 
\item take (or create) the edge in $R$ from \txtc{z} to $\txtc{z}\cdot(0,\by[k] )$. 
\end{enumerate}
By Lemma \ref{condition}, we have $\txtc{z}\cdot(0,\by[k] )=(x+1,y+0)$,  so that it remains to add the neighboring link $(\bx,\by) \stackrel{\raisebox{-3pt}{$\scriptstyle{\0}$}}{\dashrightarrow} (x+1,y+0).$ Then,  a nonempty word $w\in \Freeman^n$ is sequentially processed to build  the graph 
 $\G_{w},$
and we illustrate the algorithm on the input word $w=\0\0\1\1$.

\bigskip

\noindent
{\begin{minipage}[t]{\textwidth-2.7in}
\noindent{\em $\bullet$ Initialization}: one starts with the graph consisting of the node $(0,0)$  marked as visited. For convenience, 
the \emph{non-visited} nodes $(0,1),(1,0)$, and the links from $(0,0)$  to its neighbors are also added. This is justified by the fact 
that the algorithm applies to nonempty words.\\
\indent Since $(0,0)$ is an ancestor of all nodes, this ensures that every node has an ancestor linked with its neighbors. 
The {\em current node} is set to $(0,0)$ and this graph is called  the \emph{initial graph} $\G_{\varepsilon}$.
\end{minipage}}
{\begin{minipage}[t]{2.7in}
\mbox{}\\[-2\baselineskip]
\begin{center}
\includegraphics[width=2.5in]{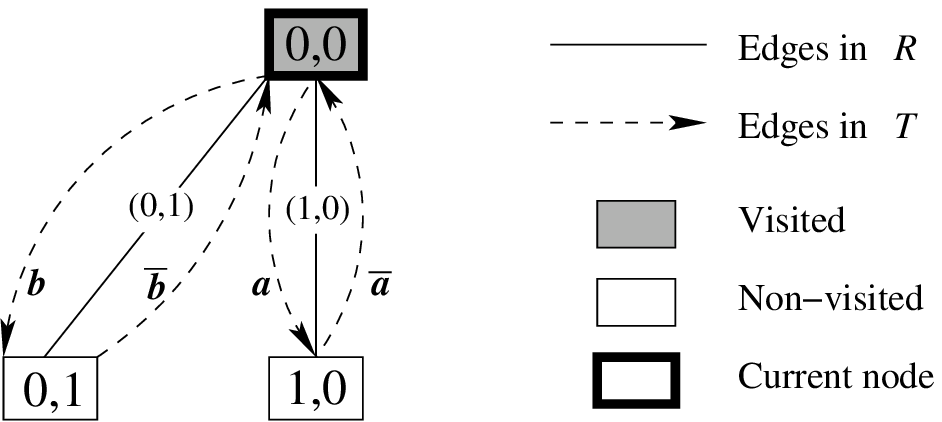}\\
\addtocounter{figure}{1}
Figure \arabic{figure}: Initial graph $\G_{\varepsilon}$.
\end{center}
\end{minipage}}


\mbox{}\\

\noindent
\begin{minipage}[t]{\textwidth-1.5in}
\noindent{\em $\bullet$ Read $w_1=\0$}: this corresponds to the translation $(0,0)+(1,0)$. A neighboring link labeled $\0$ starting from $(0,0)$ 
and leading to the node $(1,0)$ does exist, so the only thing to do is to follow this link and mark the node $(1,0)$ as visited. 
The current node is now set to $(1,0),$ and this new graph is called $\G_0$.
\end{minipage}
\begin{minipage}[t]{1.5in}
\mbox{}\\[-2\baselineskip]
\begin{center}
\includegraphics[height=0.8in]{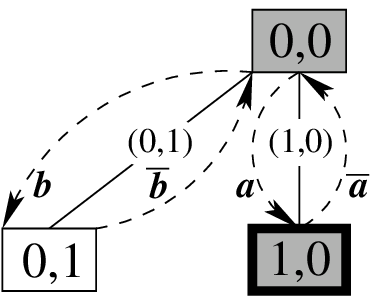}\\
\addtocounter{figure}{1}
Figure \arabic{figure}: Graph $\G_{\0}$.
\end{center}
\end{minipage}


\vspace{\baselineskip}

\noindent
\begin{minipage}[t]{\textwidth-1.5in}
\noindent{\em $\bullet$ Read $w_2 = \0$}: this time, there is no edge in  $\G_0$ labeled $\0$ starting from $(1,0)$. Using the translation rules above, we perform:
\begin{enumerate}[(1)]
\item go back to the father $f(1,0)=(0,0);$
\item follow the link $\0$ to $(1,0);$ 
\item  add  node $(2,0) \sim (1,0)\cdot (0,0) = (10,00)$. 
\end{enumerate}
Then an edge from $(1,0)$  to $(2,0)$ with label $\0$  is added to $T$. Finally the node $(2,0)$ is marked as \emph{visited}, and becomes the current node.
\end{minipage}
\begin{minipage}[t]{1.5in}
\mbox{}\\[-2\baselineskip]
\begin{center}
\includegraphics[height=1.4in]{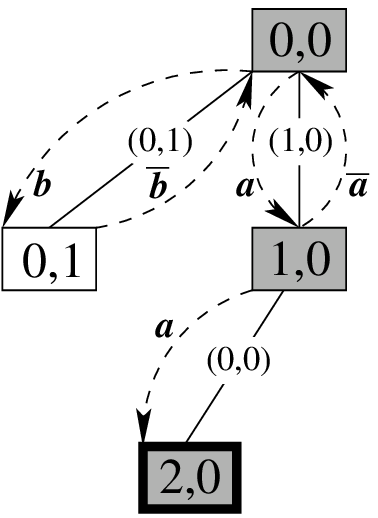}\\
\addtocounter{figure}{1}
Figure \arabic{figure}: Graph $\G_{\0\0}$.
\end{center}
\end{minipage}


\vspace{\baselineskip}

\noindent
\begin{minipage}[t]{\textwidth-1.5in}
\noindent{\em $\bullet$ Read  $w_3 = \1$}: this amounts to perform  the translation $(2,0)+(0,1)$. Since the edge to $f(2,0)$ is labeled by $(0,0)$, we know that the second coordinate of the current node $(2,0)$ is even. Therefore, $(2,1)$ and  $(2,0)$ must be siblings, that is $f((2,0)+(0,1))$ = $f((2,0))$. What we need to do then is :\smallskip\\
\makebox[20pt]{(1)} go back to the father $f(2,0)=(1,0);$\\
\makebox[20pt]{(2)} follow the edge $\1$ if it exists;\smallskip\\
Since it is does not exist,  it must be created 
to reach the node $(2,1) \sim (10,01)= (1,0)\cdot(0,1).$ 
Again an edge from $(2,0)$ to $(2,1)$ with label $\1$ is added,  $(2,1)$ is marked as \emph{visited} and is now the current node.
\end{minipage}
\begin{minipage}[t]{1.5in}
\mbox{}\\[-2\baselineskip]
\begin{center}
\includegraphics[height=1.4in]{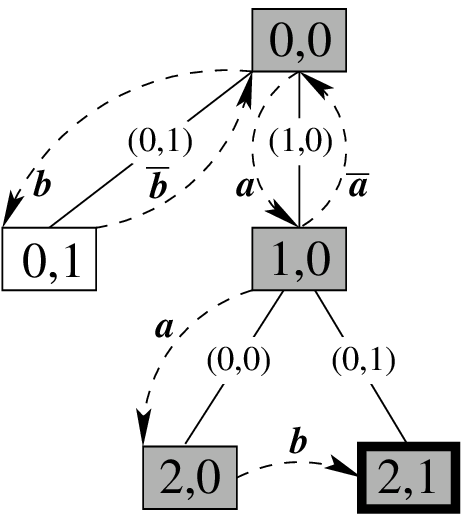}\\
\addtocounter{figure}{1}
Figure \arabic{figure}: Graph $\G_{001}$.
\end{center}
\end{minipage}


\vspace{\baselineskip}

\noindent
\begin{minipage}[t]{\textwidth-2.2in}
\noindent{\em $\bullet$ Read  $w_4 = \1$}: since  $f((2,1))$ has no neighboring link labeled by $\1$, recursion is used to find (or build if necessary)
the node corresponding to its translation by $\1$. This leads to the creation of the node $(1,1)\sim(0,0)\cdot(1,1)$  marked as \emph{non-visited}.
Then, the node $(2,2)\sim(1,1)\cdot(0,0)$  is added, marked as \emph{visited}, and becomes the current node. Note that a neighboring links between
$(1,0)$ and $(1,1)$,  $(2,1)$ and $(2,2)$ are added in order to possibly avoid searches.
\end{minipage}
\begin{minipage}[t]{2.2in}
\mbox{}\\[-2\baselineskip]
\begin{center}
\includegraphics[height=1.4in]{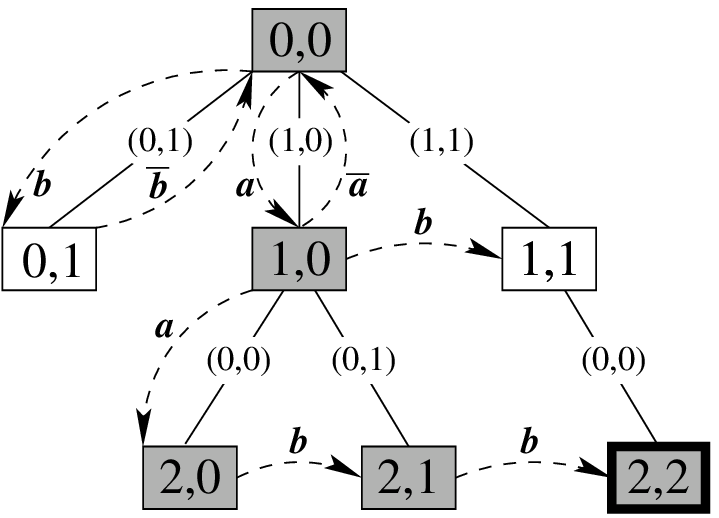}\\
\addtocounter{figure}{1}
Figure \arabic{figure}: Graph $\G_{0011}$.
\end{center}
\end{minipage}


\vspace{\baselineskip}

This algorithm is linear and we refer to  \cite {BKP} for details about the complexity analysis which is rather involved.

\section{Digital convexity{\rm \cite{blpdgci08,blpr}}}

The notion of convexity does not translate trivially, and detecting if a discrete region of the plane is convex requires a deeper analysis. 
There are several (more or less) equivalent definitions of digital
 convexity, depending on whether or not one asks the digital set to
 be connected. We say that a word $w$ is {\em digitally convex}  if it is the boundary word of a finite 4-connected subset $S$ of
 $\Z^2$ such that  $S = \ConvHull(X) \cap \Z^2$. Given such a region:
 \begin{table}[h!]
\begin{minipage}[c]{0.5 \linewidth}
\centering 
\includegraphics[height=1.25in]{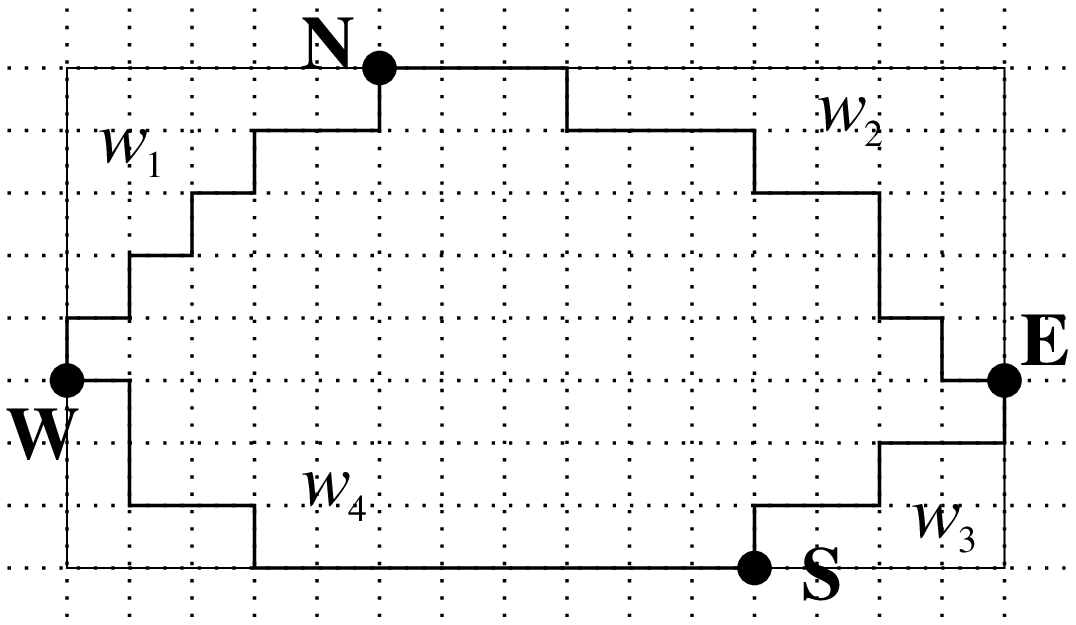}
\end{minipage}
\begin{minipage}[c]{0.5 \linewidth}
{\bf W} is the lowest on the Left side;\\
 {\bf N} is the leftmost on the Top side;\\
  {\bf E} is the highest on the Right side;\\
   {\bf S} is the rightmost on the Bottom side;\\
   So that $w\equiv w_1  w_2 w_3 w_4.$
\end{minipage}
\end{table}

\noindent
 We say that a word $w_1$ in $\Freeman^*$ is {\em NW-convex} iff there
 are no integer points between the upper convex hull of the points
 $\{\Pm{w_i}{j}\}_{j=1 \ldots |w_i|}$ and the path $w_i$. 
Then  we have 
 \[ w_i \quad \textrm{is convex} \iff \rho^{i-1}(w_i) \quad\textrm{is NW-convex}.\]
Clearly, the convexity of $w$ requires the convexity of each $w_i$ for $i=1,2,3,4$, and we have the following obvious property.  
 \noindent
Observe that if for some $i$,    $w_i$ contains more than 2 letters, that is if $\Supp(\rho^{i-1}(w_i)) \not \subseteq \{\0,\1\}$,   then $w$ is not digitally convex.  

 \begin{theorem}[\cite{blpr}]\label{main}
   A word $v$ is NW-convex iff its unique Lyndon factorization $l_1^{n_1} l_2^{n_2} \cdots l_k^{n_k}$ is such that all $l_i$ are primitive Christoffel words.
 \end{theorem}
%
%
%
%

\begin{table}[h!]
\begin{minipage}[c]{0.6 \linewidth}
For example, the Lyndon  factorization of  the word $v=\1\0\1\1\0\1\0\1\0\0\0\1\0$  is 
\[v = (\1)^1 \cdot (\0\1\1)^1 \cdot (\0\1)^2 \cdot (\0\0\0\1)^1 \cdot (\0)^1,\] 
where $\0$, $\0\1\1$, $\0\1$, $\0\0\0\1$ and $\0$ are all Christoffel words.
\end{minipage}
\begin{minipage}[c]{0.4 \linewidth}
\centering\includegraphics[height=1.1in]{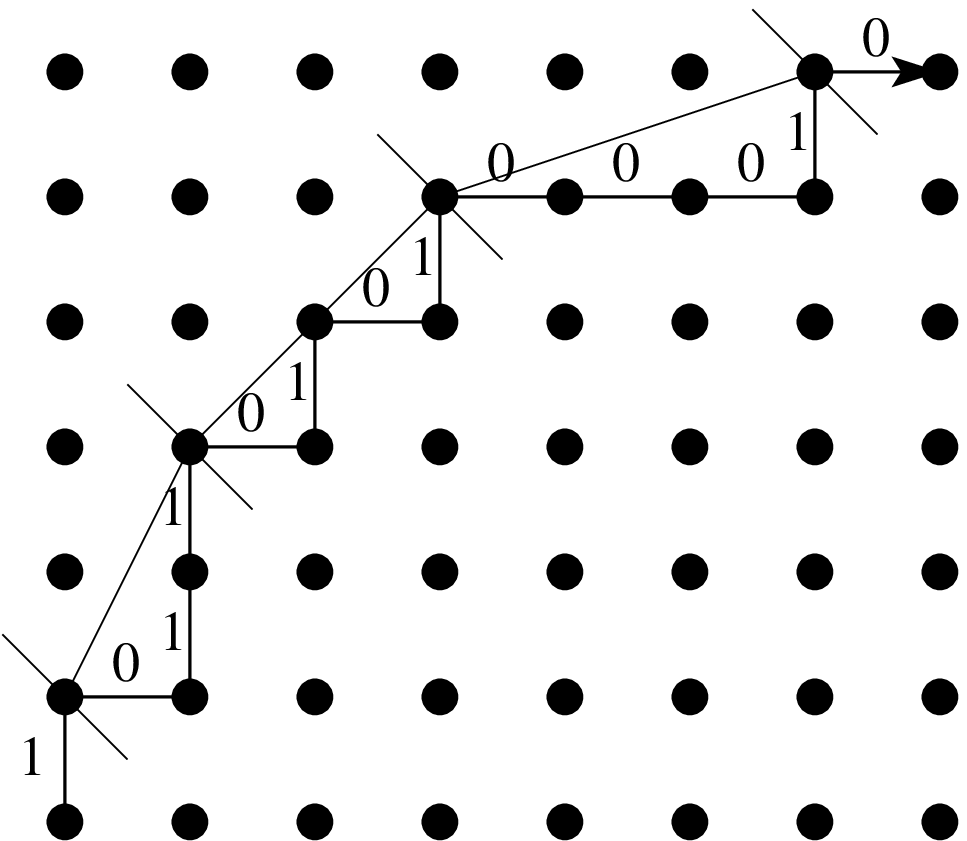}
\end{minipage}

\end{table}

This result leads to a fast optimal algorithm for checking digital convexity of a boundary word. It is based on the linear time algorithms for computing the Lyndon factorization of the contour word and for the recognition of Christoffel factors which digital line segments. By avoiding arithmetical computations the algorithm is much simpler to implement and much faster in practice (see~\cite{blpr} for more details).
\smallskip\\
It is worth noting that many results about Sturmian words have been obtained by using geometrical properties. This close relation between the two domains raises a number of combinatorial problems such as enumeration of convex words of given length~\cite{Prov11}. Another interesting property is the factorial closure of digitally convex words: while a geometrical proof is rather easy to obtain, it begs for a purely combinatorial proof. A formal language characterization is also a challenge. 

\section{Tilings}

Tilings appeared as one of the archetypes of the close relationship between art and mathematics, and are present in human history under various representations.  The beautiful book of Gr\"unbaum and Shephard  \cite{GrSh1} contains a systematic study of tilings,  presenting a number of challenging problems (see also \cite{BrMoPa05} for related work).
For instance, the problem of designing an efficient algorithm for deciding  whether a given polygon tiles the plane becomes more tractable when restricted to polyominoes, that is,  subsets of the square lattice~$\Z^2$ whose boundary is a non-crossing closed path. Indeed, while a sufficient condition is provided by the \emph{Conway criterion} in \cite{doris80}, one of its consequences is that the only objects that tile the plane by translation in two directions are generalizations of parallelograms and parallel hexagons, hexagons whose opposite sides are equal and parallel (see \cite{doris80} p. 225 for more details). One may consider these tiles are continuous deformations of either the  \emph{unit square} or the  \emph{regular hexagon}. 

\paragraph{Tiling the plane by translating a single polyomino {\rm (\cite{bpdgci06,bpfDAM})}}
Such a  polyomino is  called \emph{exact} in \cite{Wi}  and
 Beauquier and Nivat~\cite{bn} characterized them by showing that their boundary word   satisfies the equation $b(P) = X\cdot Y \cdot Z \cdot \hat{X} \cdot\hat{Y}\cdot\hat{Z}$, where at most one of the variables is empty and where $\widehat{W}=\rho^2(\widetilde{W})$. This condition is referred to as the BN-factorization.
An exact polyomino is said to be a \emph{hexagon} if none of the variables $X$, $Y$, $Z$ is empty and a \emph{square} if one of them is so. While decidability was already established in  \cite{Wi}, recently, a $\mathcal{O}(n\log^3n)$  was designed for deciding if a word $w\in \Freeman$ tiles the plane by translation \cite{bpfDAM, bp}.
 It uses several data structures that include radix-trees, for checking that $w$ is a closed non crossing path~\cite{BKP}, and suffix-trees for building a tricky algorithm for checking the BN-factorization~\cite{bpdgci06}. 

\paragraph{Square tiles} In this case there is a linear algorithm for deciding if a word is a square \cite{bpfDAM}. Moreover, it turns out that a square has at most two distinct square tilings \cite{bbgl10,bbl11}, which means that the BN-factorizations are distinct and therefore not conjugate. Motivated by the attempt of characterizing the square  tiles, the study of  equations of the form   $A\,B\,\til{A}\,\til{B}\equiv X\,Y\,\til{X}\,\til{Y}$ lead to  a subset of  solutions in bijection with the BN-factorizations.
It turns out that the solutions are strongly related to local periodicity involving palindromes and conjugate words \cite{bbgl09-2,bbgl11-2}. They also give some insight on their shape since palindromes represent symmetric sides. An interesting fact is that some infinite families of such tiles, namely the Fibonacci tiles have a fractal characteristic and are connected with some problems in number theory \cite{bblmf}, while Christoffel words yield another infinite family of squares \cite{bbgl09}. 

\begin{figure}[h!]
\centering
\includegraphics[height=.75in]{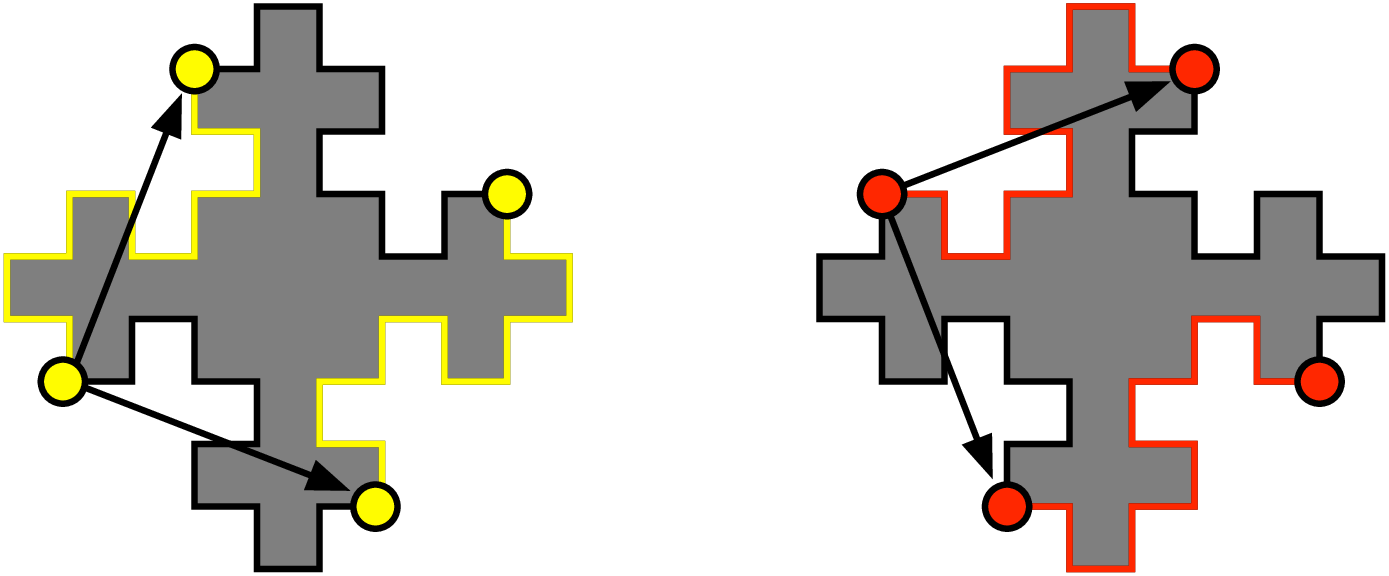}\hspace{1cm}
\includegraphics[height=1.25in]{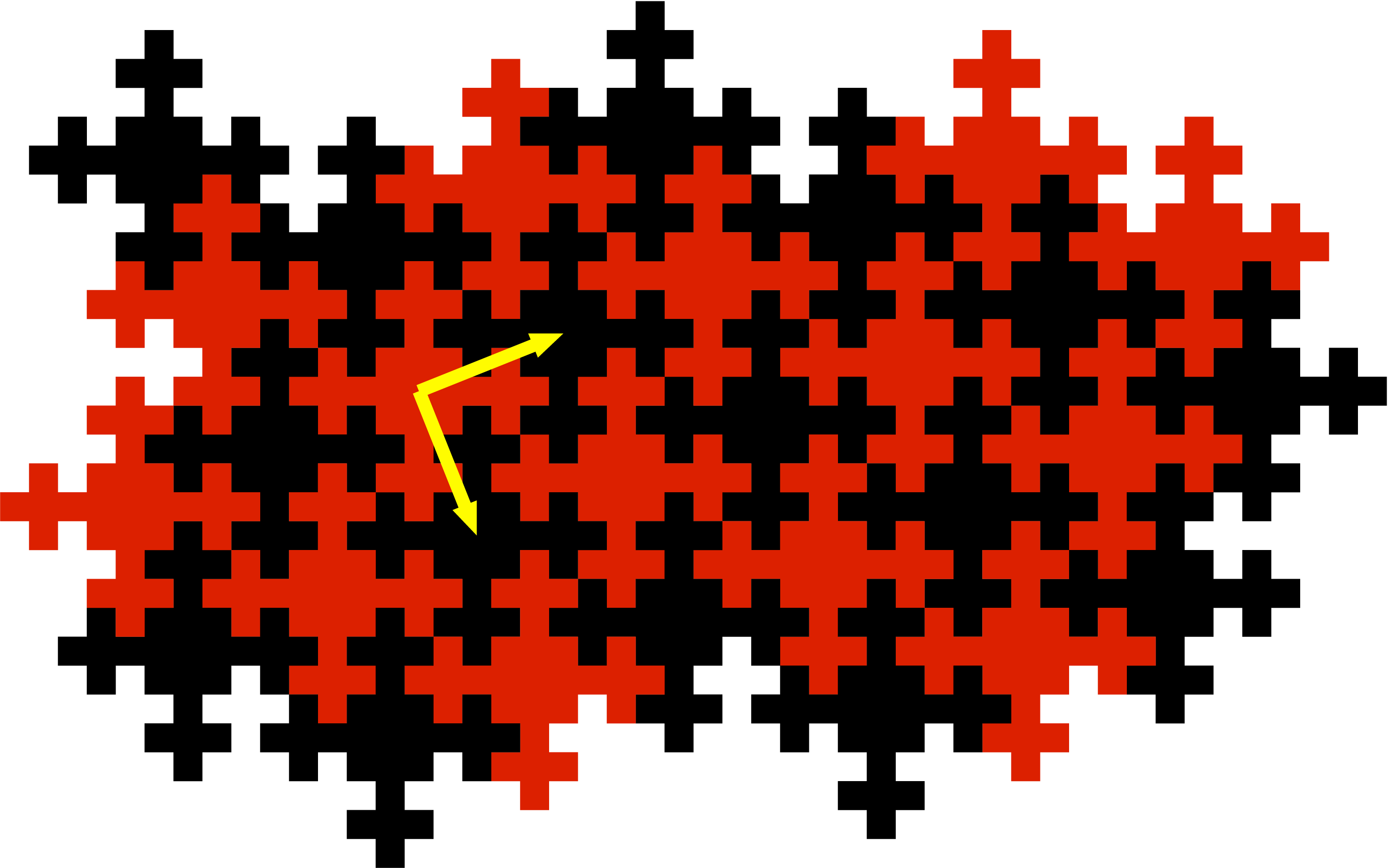}
\end{figure}

\paragraph{Remark} Again a number of problems are raised by these connections about  the enumeration and generation of tiling polyominoes. It turns out that the prototype of a square is the basic one cell polyomino, and it suffices to replace each pair of sides by homologous paths. It remains only to check that theses paths do not intersect, which is achieved in linear time thanks to the optimal algorithm we have. For the generation of double square tiles, the prototype is the cross polyomino, and their generation is a bit more involved.  
The case of hexagon grids deserves also some attention \cite{bpfDAM}.

\bibliographystyle{eptcs}
\bibliography{brlek}

\end{document}